\documentclass[10pt,conference]{IEEEtran}

\makeatletter
\def\ps@headings{%
\def\@oddhead{\mbox{}\scriptsize\rightmark \hfil \thepage}%
\def\@evenhead{\scriptsize\thepage \hfil \leftmark\mbox{}}%
\def\@oddfoot{}%
\def\@evenfoot{}}
\makeatother
\let\labelindent\relax

\usepackage{graphicx,amsmath}
\usepackage{float}
\usepackage{epsfig}
\usepackage{multirow}
\usepackage{cite,cases,url}
\usepackage{epstopdf}
\usepackage{balance}
\usepackage{enumerate}
\usepackage{graphics}
\usepackage{setspace}
\usepackage{array}
\usepackage{algorithmic}
\usepackage{algorithm}

\usepackage{enumitem}

\newcolumntype{L}{>{\centering\arraybackslash}m{5cm}}
\newcolumntype{P}{>{\centering\arraybackslash}m{2.3cm}}
\newcolumntype{M}{>{\raggedright\arraybackslash}m{2cm}}
\newcolumntype{N}{>{\raggedright\arraybackslash}m{2.5cm}}

\IEEEoverridecommandlockouts

\IEEEpubid{\begin{minipage}{\textwidth}\ \\[8pt]
  ~\copyright~Copyright 2016 Bloomberg L.P. All rights reserved.\\
  Distributed under the Creative Commons Attribution-Noncommercial-Share \\
  Alike license (CC BY-NC-SA 4.0)
\end{minipage}}

\begin{document}

\title{LTE security, protocol exploits and location tracking experimentation with low-cost software radio}

\author{\IEEEauthorblockN{Roger Piqueras Jover}
\IEEEauthorblockA{Bloomberg LP, New York, NY}
\IEEEauthorblockA{rpiquerasjov@bloomberg.net}
}

\maketitle
\begin{abstract}
The Long Term Evolution (LTE) is the latest mobile standard being implemented globally to provide connectivity and access to advanced services for personal mobile devices. Moreover, LTE networks are considered to be one of the main pillars for the deployment of Machine to Machine (M2M) communication systems and the spread of the Internet of Things (IoT). As an enabler for advanced communications services with a subscription count in the billions, security is of capital importance in LTE. Although legacy GSM (Global System for Mobile Communications) networks are known for being insecure and vulnerable to rogue base stations, LTE is assumed to guarantee confidentiality and strong authentication. However, LTE networks are vulnerable to security threats that tamper availability, privacy and authentication. This manuscript, which summarizes and expands the results presented by the author at ShmooCon 2016 \cite{jover2016lte}, investigates the insecurity rationale behind LTE protocol exploits and LTE rogue base stations based on the analysis of real LTE radio link captures from the production network. Implementation results are discussed from the actual deployment of LTE rogue base stations, IMSI catchers and exploits that can potentially block a mobile device. A previously unknown technique to potentially track the location of mobile devices as they move from cell to cell is also discussed, with mitigations being proposed.
\end{abstract}

\section{Introduction}
\label{sec:intro}
The Long Term Evolution (LTE) is the newest cellular communications standard globally deployed. Regardless of previous generations, with coexistence of different technologies for mobile access, all operators are globally converging to LTE for next generation mobile communications. With a fully redesigned PHYsical layer, built upon Orthogonal Frequency Division Multiple Access (OFDMA), LTE networks provide orders of magnitude higher rates and lower traffic latencies, combined to a strong resiliency to multipath fading and overall improved efficiency in terms of bits per second per unit of bandwidth. This highly improved Radio Access Network (RAN), is architected over the Enhanced Packet Core (EPC) network to provide connectivity to all types of mobile devices \cite{LTEBOOK}.

LTE cellular networks deliver advanced services for billions of users, beyond traditional voice and short message communications, as the cornerstone of today's digital and connected society. Moreover, mobile networks are one of the main enablers for the emergence of Machine to Machine (M2M) systems, with LTE expected to play a key role in the Inernet of Things (IoT) revolution \cite{m2m_and_LTE}. Consequently, M2M and the IoT are often analyzed as key elements within the LTE security ecosystem \cite{m2m_lte_security}.

Given the widespread usage, with a subscription count in the billions, securing the connectivity of mobile devices is of extreme importance. The first generation of mobile networks (1G) lacked of support for encryption and legacy 2G networks lack of mutual authentication and implement an outdated encryption algorithm \cite{attack_sniffing}. The wide availability of open source implementations of the GSM protocol stack has resulted in many security research projects unveiling several exploits possible on the GSM insecure radio link.

Specific efforts were thus made to ensure confidentiality and authentication in mobile networks, resulting in much stronger cryptographic algorithms and mutual authentication being explicitly implemented in both 3G and LTE. Based on this, LTE is generally considered secure given this mutual authentication and strong encryption scheme. As such, confidentiality and authentication are wrongly assumed to be sufficiently guaranteed. LTE mobile networks are still vulnerable to protocol exploits, location leaks and rogue base stations.

Based on the analysis of real LTE traffic captures obtained from live production networks in the areas of New York City and Honolulu, this manuscript discusses the insecurity rationale behind LTE protocol exploits and rogue base stations. Despite the strong cryptographic protection of user traffic and mutual authentication of LTE, a very large number of control plane (signaling) messages are exchanged over an LTE radio link in the clear regularly. Before the authentication and encryption steps of a connection are executed, a mobile device engages in a substantial conversation with *any* LTE base station (real or rogue) that advertises itself with the correct broadcast information. This broadcast information is sent in the clear and can be easily sniffed \cite{LTE_Sniffing_Roger}, allowing an adversary to easily configure and set-up a rogue access point. Real samples of LTE broadcast traffic are captured and analyzed to exemplify this and discuss potential ways this information could be leveraged with a malicious intent.

Motivated by this LTE protocol insecurity rationale, this manuscript explores different areas of mobile network security with low-cost software radio tools. Based on an open source implementation of the LTE stack, openLTE \cite{openLTE}, a series of exploits are demonstrated and implemented. Discussion is provided regarding the implementation of low-cost IMSI catchers as well as exploits that allow blocking mobile devices, which were publicly introduced for the first time in \cite{LTEpracticalattacks_Blackhat} and shortly after also discussed in \cite{jover2016lte}. Moreover, based on the analysis of real LTE traffic captures, a new location information leak in the LTE protocol is introduced, which allows potentially tracking LTE devices as they move. Potential mitigations and security discussions are provided for these exploits as well.

This manuscript is an extended report of the results presented in \cite{jover2016lte} and summarizes the author's research in LTE mobile security and protocol exploits research over the last few years. The remainder of the manuscript is organized as follows. Section \ref{sec:lte} introduces the relevant background on LTE mobile network operation. Then, Section \ref{sec:tools} and Section \ref{sec:captures} discuss the tools leveraged to analyze and implement the LTE protocol exploits herein introduced and details on the LTE traffic captures analyzed, respectively. The rationale behind the insecurity of LTE networks and discussion on LTE rogue base stations and exploits are included in Section \ref{sec:security}. Section \ref{sec:related} overviews related work on mobile network exploit analysis and security research. Finally, Section \ref{sec:conclusions} concludes the paper.

\section{LTE mobile networks}
\label{sec:lte}
LTE mobile networks split their architecture into two main sections: the Radio Access Network (RAN) and the core network, known as the Evolved Packet Core (EPC)\cite{LTEBOOK}. The RAN of an LTE network is comprised of the mobile terminals, known as User Equipment (UE), and eNodeBs, or LTE base stations. The evolution of mobile networks towards LTE highly specializes and isolates the functionality of the RAN. In current mobile deployments, the LTE RAN is able to, independently from the EPC, assign radio resources to UEs, manage their radio resource utilization, implement access control, and, leveraging the X2 interface between eNodeBs, manage mobility and handoffs.

\begin{figure}[h]
\begin{center}
\includegraphics[width=\columnwidth]{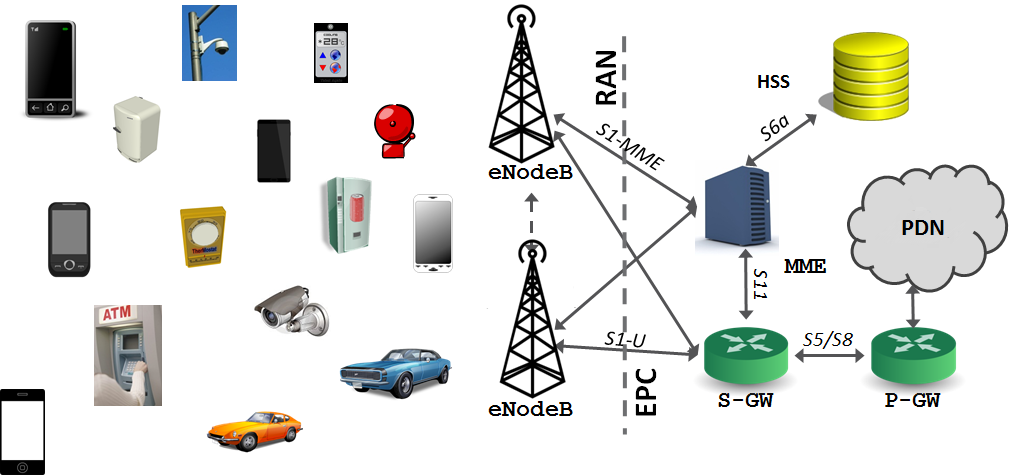}
\end{center}
\caption{LTE network architecture}
\label{fig:ltearchitecture}
\vspace{-.1in}
\end{figure}

The EPC, in turn, is in charge of establishing and managing the point-to-point connectivity between UEs and the Internet. In order to do so, the EPC leverages a series of nodes. The Serving Gateway (S-GW) and the PDN (Packet Data Network) Gateway (P-GW) are the two routing anchors for user traffic connectivity to the PDN. Once a connection is established, user data flows from the UE to the eNodeB, and is then routed over the S-GW and P--GW towards the PDN. In parallel to the routing functionality of both gateways, the Mobility Management Entity (MME) handles logistics of the bearer establishment and release, mobility management, and other network functionalities, such as authentication and access control. In order to provide security for user traffic and execute mutual authentication, the MME communicates with the Home Subscriber Server (HSS), which stores the authentication parameters, secret keys, and user account details of all the UEs. The HSS is also leveraged upon incoming connections for mobile devices, in order to address paging messages.

\begin{figure}[h]
\begin{center}
\includegraphics[width=\columnwidth]{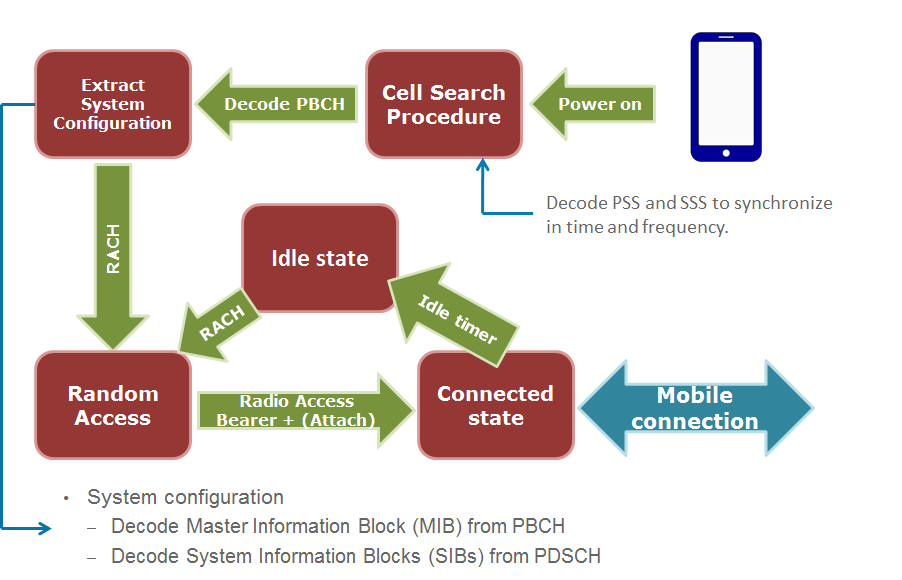}
\end{center}
\caption{LTE Cell Selection and connection}
\label{fig:lteCellSelection}
\vspace{-.1in}
\end{figure}

Any mobile device or User Equipment (UE) attempting to access the network must follow a series of steps illustrated in Figure \ref{fig:lteCellSelection}. The process is initiated by the cell selection procedure, which involves the detection and decoding of the Primary Synchronization Signal (PSS) and the Secondary Synchronization Signal (SSS). Then, the Physical Broadcast Channel (PBCH) is decoded to extract the most basic system configuration in the Master Information Block (MIB), such as the system bandwidth, which allows the other channels in the cell to be configured and operated. The remaining details of the configuration of the cell are extracted from the System Information Blocks (SIB), which are unencrypted and can be eavesdropped by a passive radio sniffer. Once at this point, the UE initiates an actual connection with the network by means of a random access procedure and establishes, via the NAS (Non-Access Stratum) Attach process, an end-to-end bearer in order to send and receive user traffic.

\begin{figure}[h]
\begin{center}
\includegraphics[width=\columnwidth]{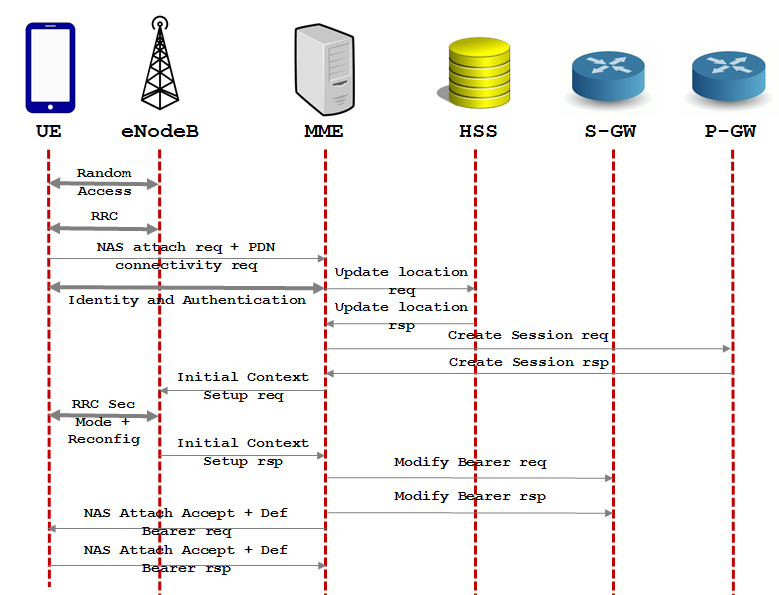}
\end{center}
\caption{LTE NAS attach procedure}
\label{fig:lteNAS}
\vspace{-.1in}
\end{figure}

Figure \ref{fig:lteNAS} illustrates the connection process through which a mobile device attaches to the network and a point-to-point IP bearer is set up to provide data connectivity. The NAS Attach procedure contains a number of steps, involving all the elements in the EPC. The Random Access procedure assigns radio resources to the UE so it can set up a Radio Resource Control (RRC) connection with the eNodeB. The next step is to execute the identity/authentication procedure between the UE and the MME, which in turn leverages the HSS to configure security attributes and encryption. Finally, the point-to-point circuit through the SGW and PGW is set up, and the RRC connection is reconfigured based on the type of QoS (Quality of Service) requested. At this point, the UE is in the \emph{connected} RRC state.

\subsection{Mobile network identifiers}

A number of identifiers are used in the operation of an LTE mobile network. The most important ones, in the context of this manuscript, are the following:

 \begin{itemize}
   \item \textbf{IMSI}: International Mobile Subscriber Identity. This is a unique identifier for the SIM card in a mobile device. The IMSI is a secret identifier that should be kept private and not transmitted in the clear as it can be used to track devices and other types of exploits \cite{engel2014ss7}.
   \item \textbf{TMSI}: Temporary Mobile Subscriber Identifier. This is the identifier used to uniquely address a given device instead of the IMSI. Once the device connects to the network for the first time, a TMSI is derived and used thereafter. The TMSI is also refreshed periodically, though not as much as it should \cite{attack_sniffing}.
   \item \textbf{MSISDN}: Mobile Subscriber ISDN Number. This is the id that identifies the user and owner of a mobile device, i.e. the user's phone number. It is mapped to the TMSI in a similar way a url is mapped to an IP address.
   \item \textbf{IMEI}: International Mobile Equipment Identifier. It uniquely identifies the hardware (i.e. smart-phone) used to connect to the network. It can be thought as a the serial number of a mobile device. The IMEI should ideally also be kept secret as, based on its value, one can easily know the type of mobile device the user has (make and model) and well as, in the case of an embedded M2M device, the software version it is running.
 \end{itemize}

\section{LTE security analysis tools}
\label{sec:tools}
Over the last few years, a number of open source projects have been developed, providing the right tools for sophisticated LTE security research. Running on off-the-shelf software radio platforms, these open source libraries provide, in some cases, the functionality of software-based eNodeB. With not too complex modifications of the code, they can easily be turned into LTE protocol analyzers, stingrays and rogue base stations, as it will be discussed throughout this manuscript.

The main LTE open source implementations being actively developed can be summarized as follows\footnote{Note that this is not an exhaustive list, but a summary of the open source tools that have been tested within the scope of this project.}:

\begin{itemize}
  \item \textbf{openLTE}\cite{openLTE}: Currently the most advanced open source implementation of the LTE stack. It provides a fully functional LTE connection, including the features of the LTE packet core network. With proper configuration, it can operate NAS protocols and provide access to the Internet for mobile devices. It implements the HSS functionality on a text file storing IMSI-key pairs.
  \item \textbf{srsLTE}\cite{srslte}: Partial implementation of the LTE stack that provides full access to PHY layer features and metrics and full access to decoded broadcast messages. The srsLTE project recently introduced srsUE, an implementation of the UE stack that allows to emulate a mobile device against an eNodeB.
  \item \textbf{gr-LTE}\cite{grLTE}: Open source LTE implementation based on gnuradio-companion, which makes it ideal for beginners to start familiarizing with the LTE protocol and signal processing steps. It mostly implements the PHY layer.
\end{itemize}

Most open source implementations can be run using standard off-the-shelf software radio, such as the USRP \cite{usrp}. This tool allows both passive and active experimentation, as it provides transmit and receive features. Passive traffic capture and eNodeB broadcast information can also be performed with much simpler platforms, such as RTL-SDR radios \cite{RTLSDR}. Note that, to abide with regulations, active experimentation (i.e. UE and eNodeB emulation) must be carried out in an RF shielded environment.

The experimentation summarized in this manuscript has been done with a Core i7 Ubuntu machine operating a USRP B210 software radio platform. All the protocol exploits described in this manuscript have been implemented by means of a customized version of openLTE, adding certain new features such as the collection of the IMSI of devices that attempt to establish a connection. All active experimentation was performed inside a Faraday cage using standard smartphones. One of the versions of eNodeB scanner was implemented using the same computing equipment and an RTL-SDR radio.

\section{LTE traffic captures}
\label{sec:captures}
Most traffic captures displayed in this manuscript were obtained with a Sanjole Wavejudge LTE sniffer and protocol analyzer \cite{sanjole}. This tool provides the means to capture and analyze traffic at the PHY layer as well as real-time capture and decoding of MAC, RRC, NAS, etc traffic. The majority of traffic capture analysis was performed with the Wavejudge software provided by the same vendor.

The LTE passive traffic captures were obtained from production networks in both the Honolulu and New York City areas, providing a real snapshot of operational LTE mobile networks in an urban environment. The traffic captures for active experimentation, i.e. IMSI catcher and device blocking, were obtained within a Faraday cage with a modified version of openLTE acting as both the eNodeB and traffic sniffer, and the author's smartphone as commmunicating endpoint.

All user traffic, despite being fully encrypted and un-decodable, was filtered out and only control plane traffic is captured and analyzed. All traffic captures included in this manuscript are edited to conceal the author's smartphone IMSI, cell IDs, mobile network operator identifiers and other sensitive data.

\section{LTE security and protocol exploits}
\label{sec:security}
This section discusses the rationale behind LTE protocol exploits based on the analysis of real LTE control plane traffic captures. Based on this, a series of security aspects and protocol exploits in the context of LTE are introduced, from leveraging the information extracted from MIB and SIB messages for rogue eNodeB configuration to denial of service threats to temporarily block mobile devices and location leaks.

\subsection{MIB and SIB message eavesdropping}
\label{sec:MIBSIB}

The MIB and SIB messages are broadcasted and mapped on the LTE frame over radio resources known a priori. Moreover, these messages are transmitted with no encryption. Therefore, any passive sniffer is able to decode them. This simplifies the initial access procedure for the UEs but could be potentially leveraged by an attacker to craft sophisticated jamming attacks, optimize the configuration of a rogue base station or tune other types of sophisticated attacks \cite{antijamming}.

\begin{figure}[h]
\begin{center}
\includegraphics[width=\columnwidth]{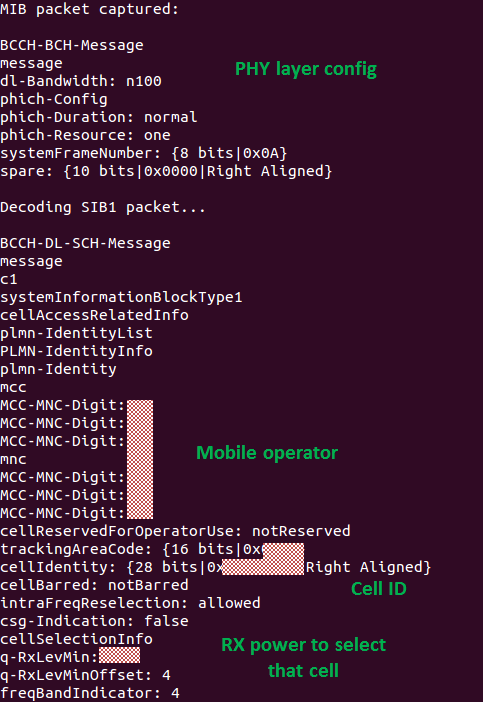}
\end{center}
\caption{Real capture of MIB and SIB1 LTE broadcast messages}
\label{fig:mibsib}
\vspace{-.1in}
\end{figure}

Figure \ref{fig:mibsib} provides an example of the contents of a MIB and SIB1 message broadcasted by a commercial eNodeB in the area of New York City. From the information extracted from these messages, an adversary can learn the mobile operator that operates that cell, the tracking area code, received power threshold to trigger a handoff to an adjacent cell and a series of configuration parameters that could be leveraged to configure a rogue base station. One of the most useful pieces of information, from a protocol exploit point of view, is the list of high priority frequencies. These can be configured on a rogue eNodeB to trigger most UEs to connect to it.

Moreover, an attacker can also extract the mapping of important control channels on the PHY layer from the SIB messages. This can be leveraged to configure a smart jammer \cite{LTE_Sniffing_Roger}.

\begin{figure}[h]
\begin{center}
\includegraphics[width=\columnwidth]{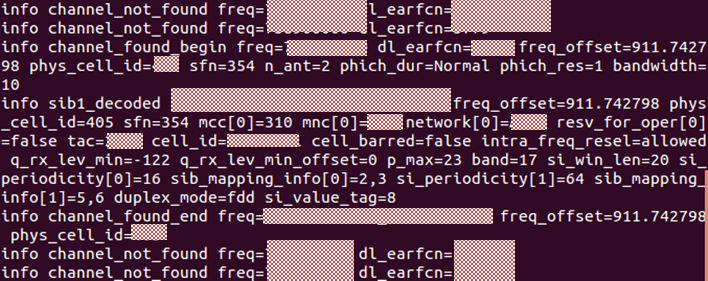}
\end{center}
\caption{Customized LTE cell scanner based on openLTE decoding and storing eNodeB MIB and SIB packets}
\label{fig:scanner}
\vspace{-.1in}
\end{figure}

OpenLTE provides a traffic log feature that can be executed to passively scan broadcast information from nearby eNodeBs. A low-cost alternative can be implemented with an RTL-SDR radio running the LTE Cell Scanner tool\cite{RTLSDR,LTEcellScanner}. The analysis of MIB and SIB traffic in this project has been carried out with a modified version of OpenLTE which automatically detects new cells, decodes MIB and SIB messages and stores the decoded information in a file. Figure \ref{fig:scanner} provides a snapshot of the cell scanner while successfully detecting a cell in the New York area and decoding the information on both MIB and SIB1 messages. The author is in the process of developing a portable LTE MIB/SIB message scanner leveraging the latest small form-factor USRP B205-mini \cite{usrp} being controlled through the USB interface of an Android smartphone. Results will be released and published once the author finds time to finsh this project and carries the scanner around for a couple of months.

\subsection{LTE insecurity rationale}
\label{sec:rationale}

The strong encryption and mutual authentication algorithms implemented in LTE networks often lead to the misunderstanding that rogue base stations are not possible in LTE. Similarly, threats such as IMSI catchers, also known as Stingrays, rogue access points and the like are generally assumed to exploit exclusively well understood vulnerabilities in latency GSM networks. However, any LTE mobile device will exchange a very large number of unencrypted and unprotected messages with any LTE base station, malicious or not, if it advertises itself with the right broadcast information. And this broadcast information can be eavesdropped by means of easily available tools, as discussed in Section \ref{sec:MIBSIB}. In order to optimize a malicious LTE access point, this can be configured with one of the high priority frequencies, which can also be extracted from unprotected broadcast messages.

\begin{figure}[h]
\begin{center}
\includegraphics[width=\columnwidth]{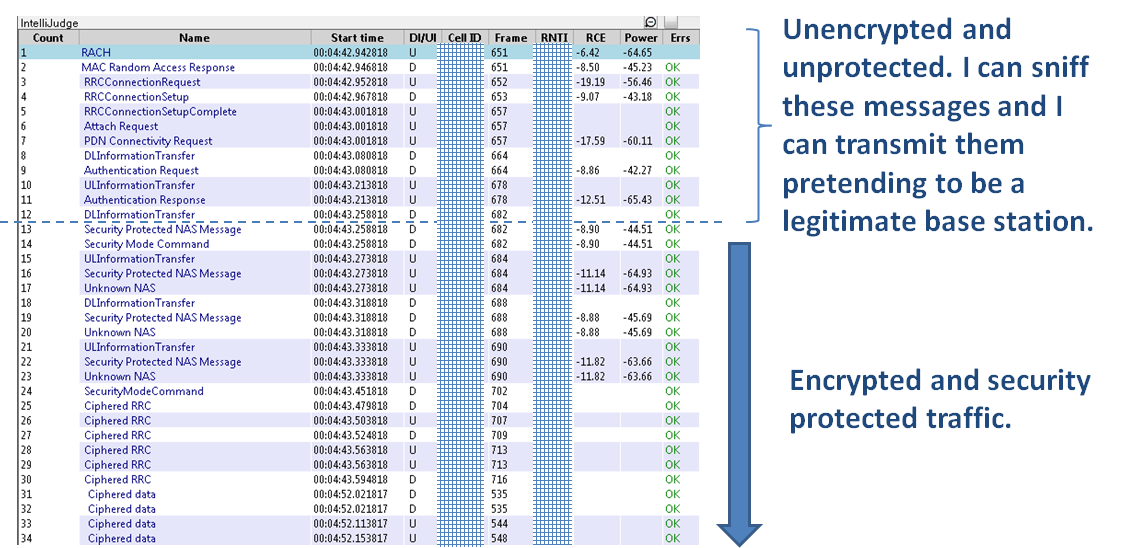}
\end{center}
\caption{Initial connection to an LTE network sniffed from a real production network in the area of Honolulu, HI}
\label{fig:rationale}
\vspace{-.1in}
\end{figure}

Figure \ref{fig:rationale} plots the actual message exchange between a mobile device and an eNodeB in order to establish a connection. The traffic was captured in a controlled environment within a Faraday cage with a single mobile device attaching to the cell. As highlighted in the figure, despite the fact that encryption is triggered upon mutual authentication, there is a large number of messages exchanged prior to the authentication step, with all this messages being sent in the clear and without integrity protection. In other words, up to the Attach Request message that a UE sends to the eNodeB, a rogue base station can impersonate a commercia eNodeB and the UE has no way to verify its legitimacy.

Aside from all the messages up to the Attach Request, there is a long list of other packets transmitted in the clear that can be eavesdropped and spoofed.

\begin{itemize}
  \item \textbf{UE measurement reports}: These reports often contain a list of nearby towers and the received power from each one of them, which can be potentially used to pinpoint the UE's location. In some rare cases, these contain explicitly the GPS coordinates of the device itself \cite{LTEpracticalattacks}.
  \item \textbf{Handover trigger messages}: As discussed later, these messages can be leveraged to track a target device as it hands from tower to tower.
  \item \textbf{Paging messages}: These messages, used by the network to locate devices in order to deliver calls and incoming connections, can be used to map a phone number to the internal id used by the network for each user.
\end{itemize}

The remainder of this section discusses how to leverage these unprotected messages to build a Stingray, temporarily block devices and potentially track users.

\subsection{LTE IMSI catcher}
\label{sec:IMSIcatcher}

Although the IMSI should always be kept private and never transmitted over the air, it is intuitive that it will be required to communicate it at least once. The very first time a mobile device is switched on and attempts to attach to the network, it only has one possible unique identifier to be used in order to authenticate with the network: the IMSI. Once the device attaches to an LTE network, a TMSI is derived and can be used thereafter to keep the IMSI secret \cite{LTEBOOK}.

\begin{figure}[h]
\begin{center}
\includegraphics[width=\columnwidth]{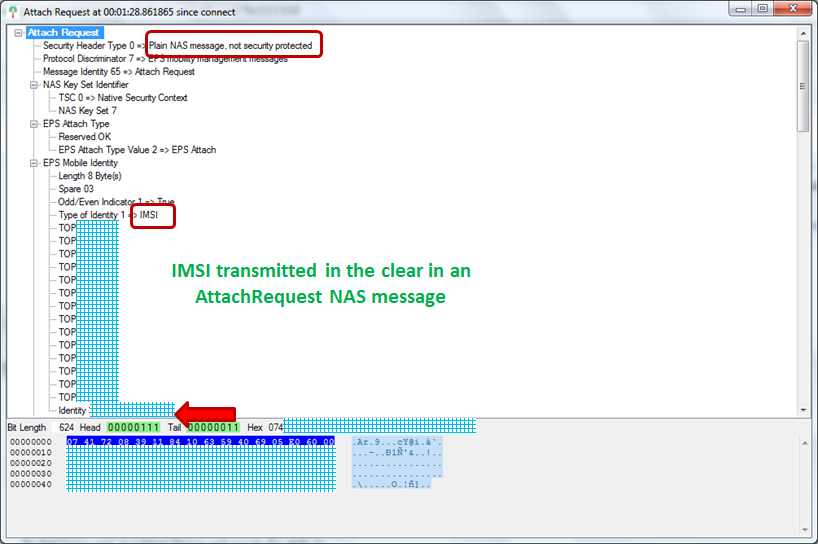}
\end{center}
\caption{IMSI of the author's smartphone being transmitted in the clear triggered by a software-radio IMSI catcher}
\label{fig:catcher}
\vspace{-.1in}
\end{figure}

As a result, regular operation of an LTE network requires that a mobile device discloses in the clear its IMSI in specific circumstances. For example, in the event that the network has never derived a TMSI for that user before or in the rare event that the TMSI has been lost. Note that the IMSI is transmitted in a message before the authentication and encryption steps of the NAS attach process. This is precisely what an IMSI catcher exploits. Figure \ref{fig:catcher} presents a real capture of an \emph{AttachRequest} message in which a mobile device is disclosing its IMSI. This was a controlled experiment with the author's smartphone in an isolated environment.

An IMSI catcher \cite{IMSIcatcher}, commonly known as Stingray, is an active radio device that impersonates a, commonly, GSM base station. In its most basic functionality, the IMSI catcher receives connection/attach request messages from all mobile devices in its vicinity. These attach messages are forced to disclose the SIM's IMSI, thus allowing the IMSI catcher to retreive the IMSI for all devices in its vicinity. Stingrays, which have been widely discussed in the generalist media recently, are known for being often used by law enforcement to track and locate suspects \cite{Stingray_TheEconomist}.

\begin{figure*}[t]
\begin{center}
\includegraphics[width=5.3in]{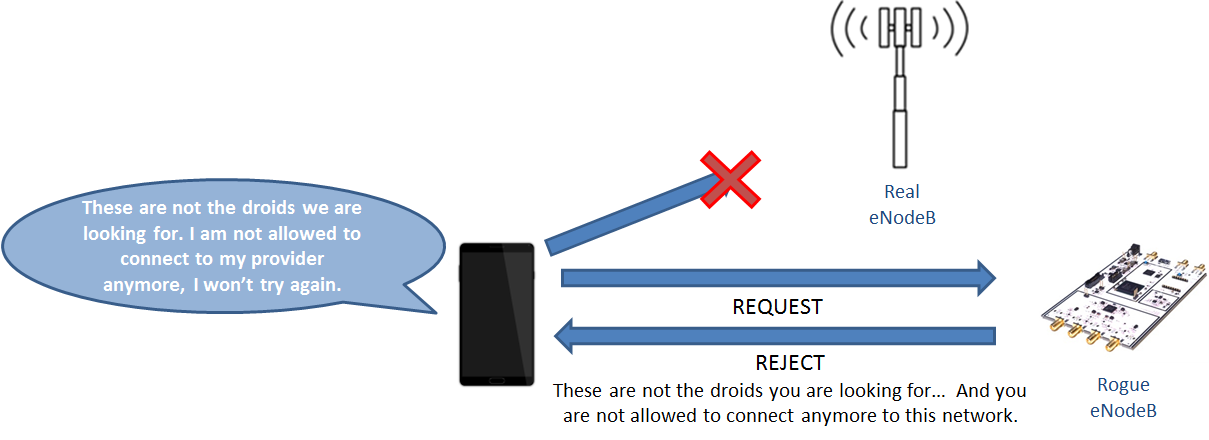}
\end{center}
\caption{Mobile device temporary block by a rogue LTE base station}
\label{fig:DoS}
\vspace{-.2in}
\end{figure*}

More advanced Stingrays actually complete the network attach process, fully impersonating a real base station. At that point, they can effectively act as a Man in the Middle (MitM) for the device's connection as long as it forwards the traffic and calls into and from the real mobile network.

Although it is well understood that a MitM Stingray is only possible in GSM, there is a general assumption that LTE IMSI catchers are not possible or, at best, highly complex and expensive \cite{piranha_LTE_IMSI_catcher}. However, a fully LTE-based IMSI catcher is possible, very simple and very cheap to implement without requiring to jam the LTE and 3G bands to downgrade the service to GSM. In the context of this project, a fully operational IMSI catcher was implemented on a USRP B210 running a modified version of openLTE. A script was added to collect and store the IMSI being disclosed by those devices within the radius of coverage of the IMSI catcher. The total budget to build the IMSI catcher was under \$2000, including the radio, antennas, GPS clock for the radio, etc. All experiments were carried out in a Faraday cage and only three IMSIs were ever collected, namely the author's smartphone IMSI and the IMSI of two LTE USB dongles. Figure \ref{fig:catcher} contains the capture of an \emph{Attach Request} message with the IMSI of the author's phone, which was captured by means of the software-radio IMSI catcher.

\subsection{Temporary blocking mobile devices}
\label{sec:block}

The basic implementation of an LTE-based IMSI catcher described in Section \ref{sec:IMSIcatcher} replies with an \emph{Attach Reject} message to the \emph{Attach Request} message, allowing the device to rapidly re-connect with the legitimate network. As with other pre-authentication messages, the \emph{Attach Reject} message is sent in the clear, which can be exploited by an attacker to temporarily block a mobile device.

3GPP defines a series of \emph{cause codes} that the network utilizes to indicate mobile devices the reason why, for example, a connection is not allowed. These codes, are defined as \emph{EMM causes} in the LTE specifications \cite{LTE_NAS}.

Some of the EMM causes indicate the device that it is not allowed to connect to the network (i.e. \emph{PLMN not allowed} EMM cause), which is a way a network provider can block a customer who engages in, for example, mobile fraud, spam, etc. Given that at the \emph{Attach Request} stage in the connection there is no authentication or encryption established yet, any mobile device implicitly trusts the EMM cause codes regardless of the legitimacy of the base station. In other words, when the \emph{Attach Reject} message is received, the base station is not authenticated yet but the UE has to obey the \emph{Attach Reject} message.

If a rogue base station replies to an incoming connection with an \emph{Attach Reject} message, it can fool the mobile device to believe that it is not allowed to connect to that given network. As a result, the device will stop attempting to connect to any base station of the provider that was being spoofed by the rogue eNodeB. Figure \ref{fig:DoS} illustrates this exploit by which a rogue base station effectively prevents any mobile device in its radio communication range from connecting to the network, resulting in a Denial of Service (DoS).

It is important to note that this is just a temporary DoS threat. By simply rebooting the device or toggling airplane mode, the device is capable again to connect to the network. Given that toggling airplane mode is commonly the first thing a user does when the cellular connection appears to be stalled, the impact of this DoS threat on smartphones would be minor. The impact could be more severe, though, in the case of an embedded device servicing an application in the context of the IoT. In the case of an embedded sensor deployed in the field with mobile network connectivity, it might be expensive and time-consuming to send a technician to reboot all the sensors being affected.

LTE mobile devices implement a timer (\emph{T3245}) which is started when an \emph{Attach Reject} message blocks the device from further attempting to connect. Upon expiration of this timer, the mobile device is allowed again to attempt to communicate and attach with the blocked network. According to the standards, this timer is configured to a value between 24 and 48 hours \cite{LTE_NAS}. Therefore, even in the context of an embedded sensor, the DoS would only be sustained for 1 to 2 days. In some applications, the impact of 24 hours of connectivity loss could be very high, specially in critical and security applications.

This same attack can be also carried over by means of replying with a reject message to the \emph{Traffic Area Update} (TAU) message. By means of configuring a rogue base station with a different Tracking Area value than the surrounding legitimate eNodeBs, one can trigger \emph{TAU} messages from the devices that attempt to connect with the rogue base station.

\begin{figure*}[t]
\begin{center}
\includegraphics[width=5.3in]{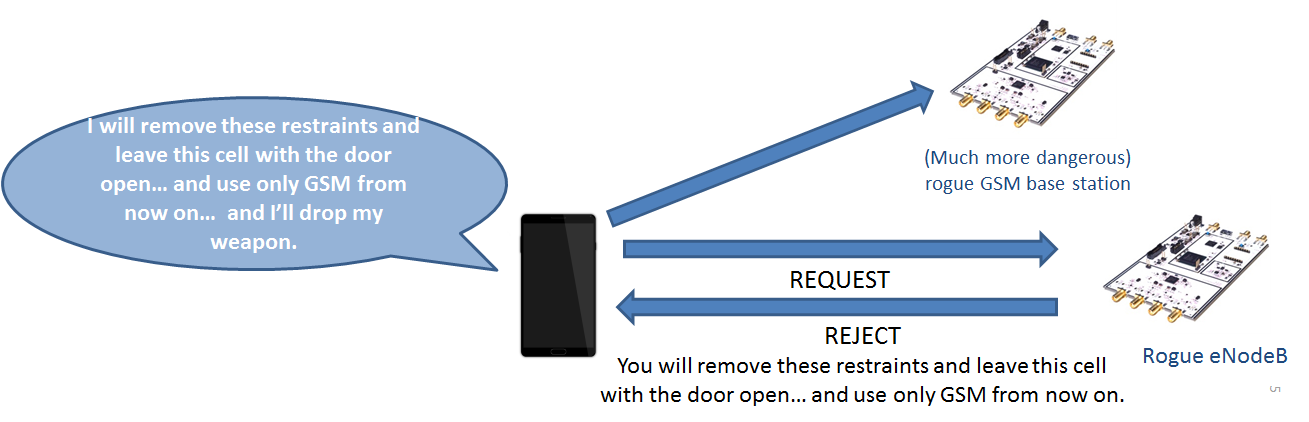}
\end{center}
\caption{Mobile device soft downgrading to GSM by rogue LTE base station}
\label{fig:downgrade}
\vspace{-.2in}
\end{figure*}

This LTE protocol exploit, which was first introduced by the authors of \cite{LTEpracticalattacks_Blackhat} and later also discussed in \cite{jover2016lte}, was implemented by means of a modified version of openLTE running on a USRP B210. All the experiments were carried in a controlled environment resulting in the blocking of the author's smartphone and 2 LTE USB dongles. No experimentation was carried over to determine the value of the timer T3245 because the author could not go about without phone for 24 hours. Excellent further analysis and results of this threat were recently presented in \cite{LTEpracticalattacks} in an outstanding paper by the authors.

\subsection{Soft downgrade to GSM}
\label{sec:GSM}

Similarly to the DoS threat discussed in Section \ref{sec:block}, an attacker can trigger a soft downgrade of the connection to GSM, known for being highly insecure \cite{attack_sniffing}. Exploiting the same \emph{TAU Reject} and \emph{Attach Reject} messages, a rogue base station can indicate a victim mobile device that it is not allowed to access 3G and LTE services on that given operator (\emph{EPS services not allowed} EMM cause code). The target mobile device will then only attempt to connect to GSM base stations, as described in Figure \ref{fig:downgrade}.

An attacker could combine this with a rogue GSM base station, which would open the doors to a full MitM threat, fully eavesdropping all mobile network traffic. This would allow the attacker to listen to phone calls, read text messages and a long list of other known GSM threats \cite{bailey2010carmen}. Moreover, by configuring the rogue base station to target a specific user, one could leverage this technology for a spearphishing threat aiming to a specific device. As the device would not loose connectivity, the user might not realize it is connected through GSM unless.

\subsection{LTE device tracking with C-RNTI}
\label{sec:RNTI}

This sub-section introduces a previously unknown LTE exploit that could potentially allow a passive adversary (i.e. only sniffing capabilities) to locate and track devices and users as they move. This location leak threat was first disclosed by the author in \cite{jover2016lte} and has already been discussed and analyzed with both GSMA and 3GPP.

The Cell Random Network Temporary Identifier (C-RNTI, RNTI for short from here on) is a PHY layer identifier unique per device within a given cell. In other words, there are no two devices within a cell with the same RNTI. But there could be two devices in adjacent cells with the same RNTI.

\begin{figure}[h]
\begin{center}
\includegraphics[width=\columnwidth]{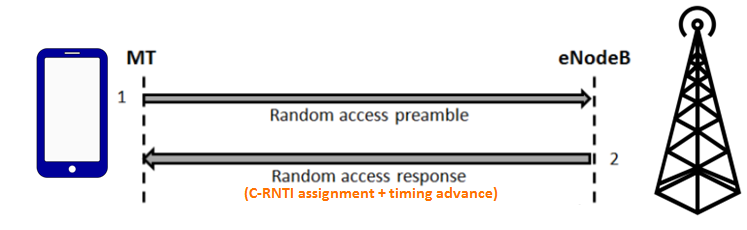}
\end{center}
\caption{RNTI assignment in the RACH LTE procedure}
\label{fig:RNTIassignment}
\vspace{-.1in}
\end{figure}

This 16-bit identifier is assigned to each mobile device during the Random Access Procedure \cite{LTEBOOK}. The eNodeB responds to the device's preamble with a MAC Random Access Response (MAC RAR) message that indicates, in the clear, the RNTI to be used by the device in that cell, as shown in Figure \ref{fig:RNTIassignment}.

\begin{figure}[h]
\begin{center}
\includegraphics[width=\columnwidth]{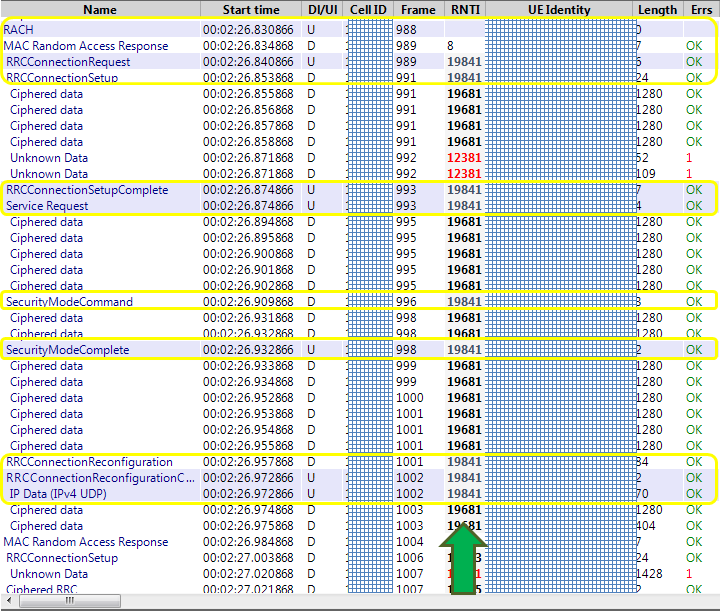}
\end{center}
\caption{Identifying the traffic of a given device with the RNTI}
\label{fig:RNTImapping}
\vspace{-.1in}
\end{figure}

Passive analysis of real LTE traffic indicates that the RNTI is included in the header (unencrypted PHY layer encapsulation) of every single packet, regardless of whether it is signaling or user traffic. This allows, as shown in Figure \ref{fig:RNTImapping}, a passive observer to easily map traffic, regardless of its encryption, to an individual device or user.

Mapping of a TMSI or MSISDN to the RNTI is trivial. By means of silent text messages or other mobile terminated traffic, an attacker can identify the current RNTI of a given device. Once this PHY layer id is known, a passive eavesdropper can know, for example, how long a given user stays at a given location. Assuming that an average smartphone will keep connecting to the network to receive email, SMSs, Whatsap messages, check for pull notifications and the like, there is going to be frequent and periodic data traffic originated or terminated at the device, allowing an eavesdropper to know the user is still there, while perhaps a partner of the eavesdropper is robbing the user's apartment. Moreover, assuming that the control plane traffic load for a mobile device is much lower than the actual user data load, the RNTI could also be leveraged to estimate the UL and DL data traffic load of a given device. This could potentially allow and adversary to identify the connectivity hotspot of an ad-hoc LTE-based network, such as the ones being considered for both first responders \cite{FirstNet} and tactical scenarios \cite{LTE_Tactical}.

Examination of the 3GPP standards \cite{LTE_RNTI} indicate that the RNTI is defined as a unique id for identifying the RRC connection and scheduling dedicated to a particular UE. Although initially assigned as a temporary id, it is promoted to a static value after connection establishment or re-establishment. There is no explicit indications in the standards regarding the requirements for the RNTI being refreshed periodically. Observations of real LTE traffic from the major operators in the United States indicate that often the RNTI remains static for long period of times. The author's smartphone was observed to maintain the same RNTI for over 4 hours.

Further analysis of LTE traffic uncovered a potential way a passive eavesdropper could track devices during mobility handover events. In LTE, handovers are network-triggered. Based on periodic measurement reports from the UEs, the eNodeBs decide whether it is necessary to hand the connection to an adjacent cell. As such, the handover is triggered by the source eNodeB through the \emph{RRC Connection Reconfiguration} message. In this packet, the source eNodeB indicates the UE what is the destination eNodeB and provides some parameters necessary for the UE to connect to the new tower. Upon reception of this message, the UE performs a random access procedure with the destination eNodeB, which assigns it a new RNTI. At this point, the handover is complete and the UE completes the RRC connection with the destination eNodeB.

During the investigation of LTE security and protocol exploits, it was discovered that, in some cases, the RNTI that was initially assigned to the UE by the destination eNodeB was always quickly updated shortly after via an \emph{RRC Connection Reconfiguration} message from the source eNodeB. Further inspection of the captures highlighted that, in the original \emph{RRC Connection Reconfiguration} message sent from the source eNodeB to initiate the handover process, a new RNTI is explicitly provided by the eNodeB in a \emph{Mobility Control Info} container \cite{LTE_RNTI}. This RNTI being explicitly provided matches the RNTI that is assigned to the UE at the destination cell via the \emph{RRC Connection Reconfiguration} message.

Based on observations of real LTE traffic, the message that triggers the handover process appears to be sent in the clear. As a result, a passive eavesdropper can potentially track a given device in a cell and, upon a handover event, follow the connection to the destination cell (indicated in the message that triggers the handover) and intercept the RNTI that is assigned to the device in this new cell. This location and handover information leak was observed to not occur always. Discussions with the GSMA security team indicated that the \emph{RRC Connection Reconfiguration} message that triggers the handover should not be transmitted in the clear.

\begin{figure}[h]
\begin{center}
\includegraphics[width=\columnwidth]{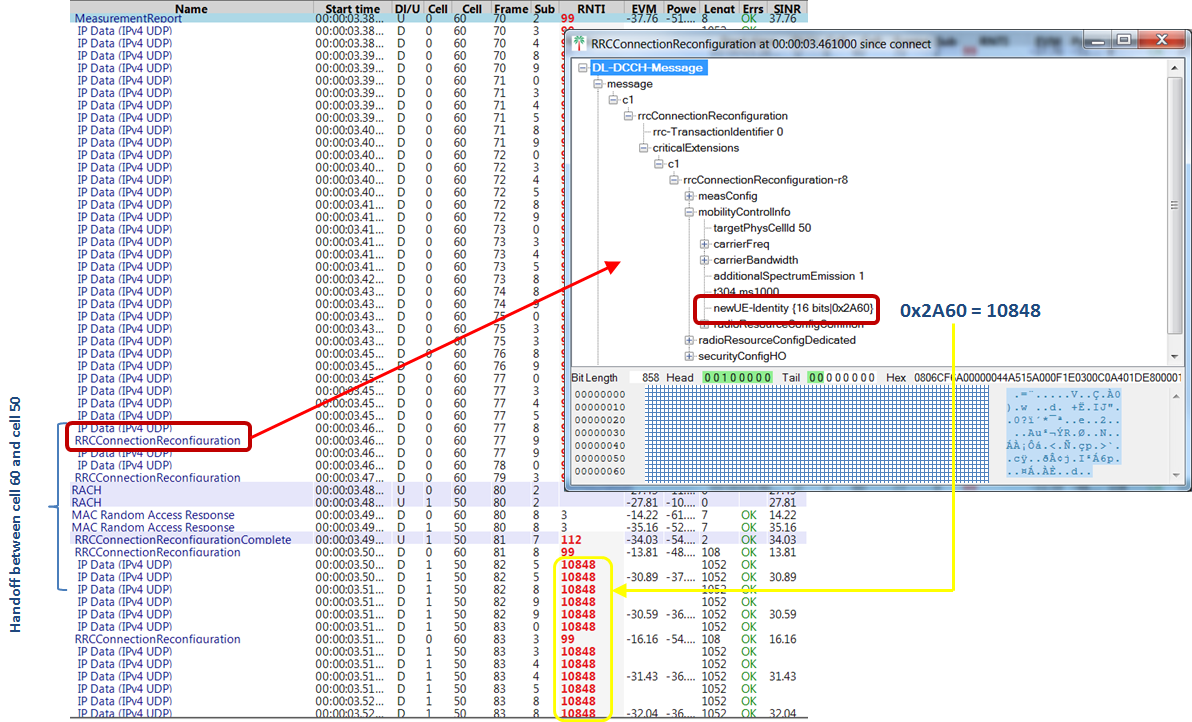}
\end{center}
\caption{RNTI device tracking during a handover process}
\label{fig:RNTItracking}
\vspace{-.1in}
\end{figure}

Figure \ref{fig:RNTItracking} presents an example of such device tracking during a handover process. In the figure, a UE with RNTI 99 is connected to cell id 60. At some point it receives an \emph{RRC Connection Reconfiguration} message triggering the connection to be handed over to cell id 50 and explicitly indicates the UE that its new RNTI at the destination cell will be 10848 (0x$2A60$). The UE performs a random access procedure and is assigned an RNTI 112 via the MAC RAR message. However, shortly after the handover, the destination eNB (cell id 50) sends an \emph{RRC Connection Reconfiguration} message that assigns the final RNTI at the destination cell, 10848. Note that, as part of the handover process, the UE still receives a couple of RRC messages from the source eNB (cell id 60) when it has already connected to the destination eNB and these are addressed to the old RNTI 99.

Although the standards do not explicitly indicate a need to refresh the RNTI periodically, this would be a strong mitigation against the threat of RNTI-based location tracking. For example, a new RNTI could be assigned each time a device transitions from idle to connected state. Also, it should be enforced that the handover trigger message is always encrypted as this message occurs after authentication and encryption set up. Nevertheless, it is important to note that these solutions might not be sufficient. An LTE analysis tool introduced in \cite{erranLTEanalytics} leverages the RNTI in order to map PHY layer measurements to a given device. In the event of an RNTI change, the authors devised an algorithm that was able to automatically map each device to its new RNTI. Keeping a fingerprint of signal measurements for each RNTI, the authors were able to track users as their RNTI changed with a precision of 98.4\%.

\section{Related work}
\label{sec:related}
The wide availability of open source tools for GSM experimentation has fueled a large array of very interesting security research on legacy networks. From the first demonstrations of GSM traffic interception and eavesdropping \cite{attack_sniffing} to DoS threats against the GSM air interface \cite{attack_RACH_GSM1}, a number of GSM exploits have been reported in the literature and security conferences. Privacy and location leaks have also been deeply investigated in the context of legacy GSM networks \cite{kune2012location}.

LTE security research has been increasingly predominant over the last couple of years, mainly in network availability related projects. For example, there has been interesting studies aiming to quantify and investigate the impact of large spikes of traffic load originated from M2M systems against the LTE infrastructure \cite{m2m_phy_impact_modeling}. Also in the context of M2M, several studies have focused on the control plane signaling impact of the IoT against the LTE mobile core \cite{jaber2014cellular,M2MscalabilityJermyn}.

Applied LTE security research and protocol exploit experimentation have been close to non-existent over the last few years. However, the recent availability of open source tools for LTE experimentation have provided the means for very interesting security research work. For example, some recent studies aimed at analyzing and evaluating sophisticated jamming threats against LTE networks \cite{lichtman2013security,lichtman2013vulnerability}. Also leveraging LTE open source tools, the authors of \cite{LTEpracticalattacks} were the first to publicly disclose the implementation and analysis of the device blocking and soft downgrade to GSM exploits, which were also implemented in this manuscript. The same authors are responsible for some other excellent mobile protocol exploit experimentation, such as a study on intercepting phone calls and text messages in GSM networks \cite{golde2013letmeanswer} and mobile phone baseband fuzzing \cite{mulliner2011smsfuzzing}.

\section{Conclusions}
\label{sec:conclusions}
This manuscript summarizes the experimentation and results of analyzing the security of next generation LTE networks with low-cost software-radio tools. Based on tools built upon the openLTE implementation of the LTE stack, the rationale behind a number of LTE protocol exploits is defined. Despite mutual authentication and strong encryption, there is still a large number of packets being exchanged, in the clear, between a UE and an eNodeB prior to these security functions being executed. A mobile device implicitly trusts these messages as long as they originate from an eNodeB broadcasting the right information, which opens the doors to a series of LTE rogue base station threats.

Basic software-radio tools can be used to scan the LTE broadcast channels so that a rogue access point can be correctly configured. Once mobile devices attempt to attach to this malicious eNodeB, different exploits are possible by leveraging the unencrypted pre-authentication messages. This manuscript provides details on the implementation of an LTE-based IMSI catcher using off the shelf hardware with a budget under \$2000.

Moreover, we investigate and implement DoS threats that block mobile devices and silently downgrade them to an insecure GSM connection. Such results are achivable by leveraging the EMM cause codes within the \emph{Attach Reject} messages the rogue eNodeB transmits. Finally, a previously unknown LTE location leak is introduced and analyzed. By means of monitoring the RNTI PHY layer id, one could potentially track and follow a mobile user as it hands from tower to tower.

The growing number of LTE open source implementations is lowering the bar for software-radio experimentation and analysis of mobile protocols. Thanks to the efforts of the open source community, such tools are fueling sophisticated research studies aimed at improving the security of mobile networks.

\section*{Acknowledgment}
The author would like to extend his most sincere gratitude to Sanjole for providing some of the LTE traffic captures used in this manuscript.

\balance

\bibliographystyle{IEEEtran}
\bibliography{exploits_bib}

\begin{thebibliography}{10}
\providecommand{\url}[1]{#1}
\csname url@samestyle\endcsname
\providecommand{\newblock}{\relax}
\providecommand{\bibinfo}[2]{#2}
\providecommand{\BIBentrySTDinterwordspacing}{\spaceskip=0pt\relax}
\providecommand{\BIBentryALTinterwordstretchfactor}{4}
\providecommand{\BIBentryALTinterwordspacing}{\spaceskip=\fontdimen2\font plus
\BIBentryALTinterwordstretchfactor\fontdimen3\font minus
  \fontdimen4\font\relax}
\providecommand{\BIBforeignlanguage}[2]{{%
\expandafter\ifx\csname l@#1\endcsname\relax
\typeout{** WARNING: IEEEtran.bst: No hyphenation pattern has been}%
\typeout{** loaded for the language `#1'. Using the pattern for}%
\typeout{** the default language instead.}%
\else
\language=\csname l@#1\endcsname
\fi
#2}}
\providecommand{\BIBdecl}{\relax}
\BIBdecl

\bibitem{jover2016lte}
R.~P. Jover, ``{LTE security and protocol exploits},'' \emph{{Shmoocon 2016}},
  January 2016.

\bibitem{LTEBOOK}
S.~Sesia, M.~Baker, and I.~Toufik, \emph{LTE, The UMTS Long Term Evolution:
  From Theory to Practice}.\hskip 1em plus 0.5em minus 0.4em\relax Wiley, 2009.

\bibitem{m2m_and_LTE}
D.~Lewis, ``{Closing in on the Future With 4G LTE and M2M},'' {Verizon Wireless
  News Center}, September 2012, \url{http://goo.gl/ZVf7Pd}.

\bibitem{m2m_lte_security}
A.~Prasad, ``{3GPP SAE-LTE Security},'' in \emph{{NIKSUN WWSMC}}, July 2011.

\bibitem{attack_sniffing}
K.~Nohl and S.~Munaut, ``Wideband {GSM} sniffing,'' in \emph{In 27th Chaos
  Communication Congress}, 2010, \url{http://goo.gl/wT5tz}.

\bibitem{LTE_Sniffing_Roger}
M.~Lichtman, R.~Piqueras~Jover, M.~Labib, R.~Rao, V.~Marojevic, and J.~H. Reed,
  ``{LTE/LTE-A Jamming, Spoofing and Sniffing: Threat Assessment and
  Mitigation},'' \emph{Communications Magazine, IEEE}, vol.~54, no.~4, 2016.

\bibitem{openLTE}
``{OpenLTE - An open source 3GPP LTE implementation},''
  \url{http://openlte.sourceforge.net/}.

\bibitem{LTEpracticalattacks_Blackhat}
A.~Shaik, R.~Borgaonkar, N.~Asokan, V.~Niemi, and J.-P. Seifert, ``{LTE and
  IMSI catcher myths},'' \emph{Proc. of BlackHat Europe}, 2015.

\bibitem{engel2014ss7}
T.~Engel, ``{SS7: Locate. Track. Manipulate},'' in \emph{FTP: http://events.
  ccc. de/congress/2014/Fahrplan/system/attachments/2553/or
  iginal/31c3-ss7-locate-track-manipulate. pdf}, 2014.

\bibitem{srslte}
I.~Gomez-Miguelez, A.~Garcia-Saavedra, P.~D. Sutton, P.~Serrano, C.~Cano, and
  D.~J. Leith, ``{srsLTE: An Open-Source Platform for LTE Evolution and
  Experimentation},'' \emph{arXiv preprint arXiv:1602.04629}, 2016.

\bibitem{grLTE}
``{gr-LTE - Gnuradio LTE cellular receiver},''
  \url{https://github.com/kit-cel/gr-lte}.

\bibitem{usrp}
{Ettus Research}, ``{USRP},'' \url{http://www.ettus.com/}.

\bibitem{RTLSDR}
``{RTL-SDR},'' \url{http://www.rtl-sdr.com/}.

\bibitem{sanjole}
{Sanjole}, ``{WaveJudge 4900A LTE analyzer},'' \url{http://goo.gl/ZG6CCX}.

\bibitem{antijamming}
R.~{Piqueras Jover}, J.~Lackey, and A.~Raghavan, ``{Enhancing the security of
  LTE networks against jamming attacks},'' February 2013, Under submission.

\bibitem{LTEcellScanner}
``{Evrytania LTE tools - LTE Cell Scanner},''
  \url{http://www.evrytania.com/lte-tools}.

\bibitem{LTEpracticalattacks}
A.~Shaik, R.~Borgaonkar, N.~Asokan, V.~Niemi, and J.-P. Seifert, ``{Practical
  attacks against privacy and availability in 4G/LTE mobile communication
  systems},'' in \emph{{Proceedings of the 23rd Annual Network and Distributed
  System Security Symposium (NDSS 2016)}}, 2016.

\bibitem{IMSIcatcher}
D.~Strobel, ``{IMSI catcher},'' \emph{Chair for Communication Security,
  Ruhr-Universit{\"a}t Bochum}, p.~14, 2007.

\bibitem{Stingray_TheEconomist}
``{The StingRay’s tale},'' in \emph{{The Economist}}, January 2016,
  \url{http://goo.gl/wqwL5e}.

\bibitem{piranha_LTE_IMSI_catcher}
{Insider Surveillance}, ``{Rayzone: Piranha LTE IMSI Catcher},'' Tech. Rep.,
  June 2015, \url{https://goo.gl/O2tO2o}.

\bibitem{LTE_NAS}
{Universal Mobile Telecommunications System (UMTS) - LTE},
  ``{Non-Access-Stratum (NAS) protocol for Evolved Packet System (EPS) - Stage
  3. 3GPP TS 24.301},'' vol. v9.11.0, 2013.

\bibitem{bailey2010carmen}
D.~Bailey and N.~DePetrillo, ``{The Carmen Sandiego Project},'' \emph{Proc. of
  BlackHat (Las Vegas, NV, USA, 2010)}, 2010.

\bibitem{FirstNet}
``{Nationwide Public Safety Broadband Network},'' {US Department of Homeland
  Security: Office of Emergency Communications}, June 2012,
  \url{http://goo.gl/AoF41}.

\bibitem{LTE_Tactical}
A.~Thompson, ``{Army examines feasibility of integrating 4G LTE with tactical
  network},'' {The Official Homepage of the United States Army}, 2012,
  \url{http://goo.gl/F60YNA}.

\bibitem{LTE_RNTI}
{LTE; Evolved Universal Terrestrial Radio Access (E-UTRA)}, ``{Long Term
  Evolution (LTE) physical layer; Overall description. 3GPP TS 36.300},'' vol.
  v8.11.0, 2009.

\bibitem{erranLTEanalytics}
S.~Kumar, E.~Hamed, D.~Katabi, and L.~Erran~Li, ``{LTE radio analytics made
  easy and accessible},'' in \emph{ACM SIGCOMM Computer Communication Review},
  vol.~44, no.~4.\hskip 1em plus 0.5em minus 0.4em\relax ACM, 2014, pp.
  211--222.

\bibitem{attack_RACH_GSM1}
D.~Spaar, ``{A practical DoS attack to the {GSM} network},'' in \emph{In
  DeepSec}, 2009, \url{http://tinyurl.com/7vtdoj5}.

\bibitem{kune2012location}
D.~F. Kune, J.~Koelndorfer, N.~Hopper, and Y.~Kim, ``{Location leaks on the GSM
  Air Interface},'' \emph{ISOC NDSS (Feb 2012)}, 2012.

\bibitem{m2m_phy_impact_modeling}
C.~Ide, B.~Dusza, M.~Putzke, C.~Muller, and C.~Wietfeld, ``{Influence of M2M
  communication on the physical resource utilization of LTE},'' in
  \emph{Wireless Telecommunications Symposium (WTS), 2012}.\hskip 1em plus
  0.5em minus 0.4em\relax IEEE, 2012, pp. 1--6.

\bibitem{jaber2014cellular}
M.~Jaber, N.~Kouzayha, Z.~Dawy, and A.~Kayssi, ``{On cellular network planning
  and operation with M2M signalling and security considerations},'' in
  \emph{Communications Workshops (ICC), 2014 IEEE International Conference
  on}.\hskip 1em plus 0.5em minus 0.4em\relax IEEE, 2014, pp. 429--434.

\bibitem{M2MscalabilityJermyn}
J.~Jermyn, R.~P. Jover, I.~Murynets, M.~Istomin, and S.~Stolfo, ``{Scalability
  of Machine to Machine systems and the Internet of Things on LTE mobile
  networks},'' in \emph{World of Wireless, Mobile and Multimedia Networks
  (WoWMoM), 2015 IEEE 16th International Symposium on a}.\hskip 1em plus 0.5em
  minus 0.4em\relax IEEE, 2015, pp. 1--9.

\bibitem{lichtman2013security}
T.~C. Clancy, M.~Norton, and M.~Lichtman, ``Security challenges with
  lte-advanced systems and military spectrum,'' in \emph{Military
  Communications Conference, MILCOM 2013-2013 IEEE}.\hskip 1em plus 0.5em minus
  0.4em\relax IEEE, 2013, pp. 375--381.

\bibitem{lichtman2013vulnerability}
M.~Lichtman, J.~H. Reed, T.~C. Clancy, and M.~Norton, ``{Vulnerability of LTE
  to hostile interference},'' in \emph{Global Conference on Signal and
  Information Processing (GlobalSIP), 2013 IEEE}.\hskip 1em plus 0.5em minus
  0.4em\relax IEEE, 2013, pp. 285--288.

\bibitem{golde2013letmeanswer}
N.~Golde, K.~Redon, and J.-P. Seifert, ``Let me answer that for you: exploiting
  broadcast information in cellular networks,'' in \emph{Proceedings of the
  22nd USENIX conference on Security}.\hskip 1em plus 0.5em minus 0.4em\relax
  USENIX Association, 2013, pp. 33--48.

\bibitem{mulliner2011smsfuzzing}
C.~Mulliner, N.~Golde, and J.-P. Seifert, ``{SMS of Death: From Analyzing to
  Attacking Mobile Phones on a Large Scale.}'' in \emph{USENIX Security
  Symposium}, 2011.

\end{thebibliography}

\end{document}